\documentclass{ws-p10x7}
\def\be{\begin{equation}}
\def\ee{\end{equation}}
\def\bea{\begin{eqnarray}}
\def\eea{\end{eqnarray}}
\def\nn{\nonumber}
\def\dahad{\Delta \alpha_{had}^{(5)} (m_Z)}
\def\ov{\overline}
\def\s2t{\sin^2 \theta_{eff}^{lept}} 
\begin{document}

\title{The Higgs puzzle: experiment and theory}

\author{Fabio Zwirner}

\address{Dipartimento di Fisica, Universit\`a di 
Roma `La Sapienza', and INFN, Sezione di Roma, 
Piazzale Aldo Moro 2,  I-00185 Roma, ITALY \\
E-mail: fabio.zwirner@roma1.infn.it}

\twocolumn[\maketitle\abstract{
The present experimental and theoretical knowledge of the 
physics of electroweak symmetry breaking is reviewed. Data 
still favor a light Higgs boson, of a kind that can be 
comfortably accommodated in the Standard Model or in its Minimal 
Supersymmetric extension, but exhibit a non--trivial structure
that leaves some open questions. The available experimental 
information may still be reconciled with the absence of a 
light Higgs boson, but the price to pay looks excessive. 
Recent theoretical ideas, linking the weak scale with the 
size of possible extra spatial dimensions, are briefly 
mentioned. It is stressed once more that experiments at 
high--energy colliders, such as the Tevatron and the LHC, 
are the crucial tool for eventually solving the Higgs puzzle.
}]

Rome is a city so full of religious symbols that it 
provides some inspiration on how to organize a talk 
on the physics of electroweak symmetry breaking, where 
firm experimental and theoretical results are mixed, 
so far, with a certain amount of beliefs. 

\section{The Standard Model \\ (The Orthodoxy)}

The obvious starting point for any discussion of the 
Higgs puzzle ($\equiv$~`what is the physics of electroweak
symmetry breaking?') is the Standard Model (SM), by 
now firmly established as the renormalizable quantum 
field theory of strong and electroweak interactions 
at presently accessible energies: in the spirit of 
the preface, it can be called `The Orthodoxy'. The only 
SM ingredient still escaping experimental detection, in 
a theoretical construction that works incredibly well, 
is the Higgs~\cite{higgs} boson, $H$. Its properties 
are controlled by some well--known parameters of the 
fermion and gauge sectors (including the Fermi 
constant $G_F$, which sets the value of the weak 
scale) plus an independent one, the Higgs mass $m_H$.

The elementary complex spin--0 field $\phi$, an $SU(2)_L$
doublet of weak hypercharge $Y=+1/2$ (in the normalization
where $Q=T_{3L}+Y$), is by now considered to be an essential 
part of what we call the SM. Indeed, it plays a fundamental
role in the description of two symmetry--breaking phenomena.
The first is the spontaneous breaking of the $SU(2)_L \times
U(1)_Y$ gauge symmetry down to the $U(1)_Q$ of QED,
described by the following part of the SM Lagrangian:
\be
\label{calLS}
{\cal L}_S \! = \! ( D^\mu \phi )^\dagger 
( D_\mu \phi ) -  \mu^2 \, \phi^\dagger 
\phi - \lambda ( \phi^\dagger \phi )^2 
\, .
\ee
The second is the explicit breaking of the global flavor
symmetry that is present if only gauge interactions are
switched on. This is realized by the Yukawa part of the
SM Lagrangian,
\bea
{\cal L}_{Y} &  = & 
h^U \overline{q_L} u_R \widetilde{\phi}
+ 
h^D \overline{q_L} d_R \phi
\nn \\[4pt]
&&
+
h^E \overline{l_L} e_R \phi
+ 
{\rm h.c.} \, ,
\label{calLY}
\eea
where $\widetilde{\phi} = i \sigma^2 \phi^*$, $h^{U,D,E}$ are $3
\times 3$ complex matrices and generation indices have been omitted. 
Over the years, our confidence in this description has been 
progressively reinforced by increasingly precise tests of both these 
symmetry--breaking phenomena (here the focus will be on gauge symmetry 
breaking, since the theoretical aspects of flavor symmetry breaking 
are discussed in another talk at this Conference~\cite{barbieri}). 
However, the ultimate, crucial test of the SM remains the direct 
search for the Higgs particle, by far the most important experimental 
enterprise in today's particle physics.

\subsection{Direct searches for the SM Higgs}
\label{subsec:direct}

The experimental status of the searches for the SM Higgs
particle is reviewed in detail in another talk at this 
Conference~\cite{hanson}. Here I will just summarize the
present situation from the preliminary LEP--combined 
results~\cite{lhwg} released for the conferences of 
Summer 2001:
\begin{itemize}
\item
The data still show an excess, at the level of $2.1 \sigma$,
over the expected SM background, mainly due to ALEPH data
and to the four--jet final state (to be compared with the
$2.9 \sigma$ excess in the preliminary data of November 2000).
\item
The maximum likelihood occurs at $m_H = 115.6 \; {\rm GeV}$,
with $3.5 \%$ probability of a fluctuation of the SM background.
\item
The lower bound on the Higgs mass is $m_H > 114.1  {\; \rm GeV}$ 
at $95 \%$ c.l., to be compared with an expected bound of 115.4~GeV.
\item
Three out of the four LEP experiments have not yet released their
final results at the time of this Conference: the final combination 
of the LEP results is expected for the end of 2001.
\end{itemize}

\subsection{SM fits to the Higgs mass}

\begin{figure}[ht]
\epsfxsize200pt
\figurebox{120pt}{160pt}{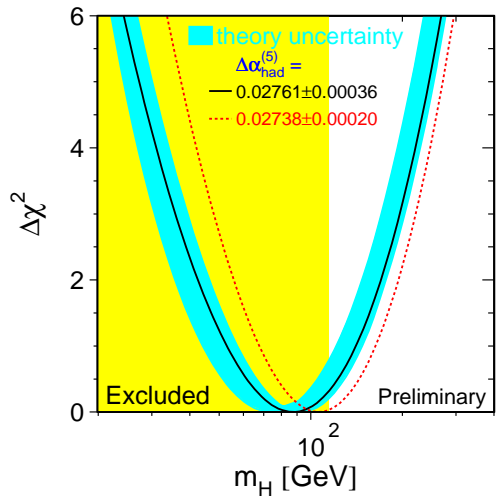}
\vspace*{-0.2cm}
\caption{$\Delta \chi^2$ as a function of $m_H$ from the 
global fit to the SM. The region excluded by the direct 
searches at LEP is also shown.}
\label{fig:blueband}
\vspace*{-0.2cm}
\end{figure}

Besides direct searches, additional information on the SM 
Higgs boson come from the fits to $m_H$ based on electroweak
precision data, whose experimental aspects are discussed 
in another talk at this Conference~\cite{drees}. A popular 
summary of the available information~\cite{lepewwg} is the 
famous `blueband' plot of the LEP Electroweak Working Group, 
displayed in Fig.~\ref{fig:blueband}: it gives the 
$\Delta \chi^2 = \chi^2 - \chi^2_{min}$ of the global SM 
fit to electroweak precision data as a function of $m_H$.
As evident from Fig.~\ref{fig:blueband}, the fit clearly 
favours a light Higgs. The default fit, represented by 
the solid curve in Fig.~\ref{fig:blueband}, gives $m_H = 
88^{+53}_{-35}$~GeV and $m_H < 196$~GeV at $95 \%$~c.l.. 
An alternative fit (see below for an explanation), 
represented by the dashed line, gives $m_H < 222$~GeV 
at $95 \%$~c.l. and a slightly higher central value. 
The band represents a (debatable) estimate of the 
theoretical uncertainty. Notice that the fit does not 
include the information coming from direct searches. 
Notice also that, in both fits, more than half of the 
$\chi^2$ curve falls in the shaded region, excluded at 
$95 \%$~c.l. by direct searches.

Given the importance of the issue, it is worth examining in more 
detail how the preference for a light SM Higgs arises. For given 
values of the remaining SM input parameters, precise electroweak 
data combined with updated theoretical calculations give logarithmic 
sensitivity to $m_H$, mostly via two pseudo--observables: the
leptonic effective electroweak mixing angle, $\s2t = (1 - v_l / 
a_l) / 4$, and the mass of the $W$ boson, $m_W$. There are still
small theoretical uncertainties in the evaluation of radiative
corrections, in principle reducible by more refined calculations. 
A recent progress along these lines is the calculation~\cite{fhww} 
of the complete fermionic two--loop contribution to $m_W$, but 
other calculations of the same order are still missing.
Larger uncertainties come from non--negligible errors in other 
parameters entering the fit. An important one is the hadronic 
contribution to the running of the electromagnetic coupling 
constant, $\Delta \alpha_{had}^{(5)}$, as extracted from a
dispersion integral over a parametrization of the measured
cross--section for $e^+ e^- \rightarrow {\rm hadrons}$,
including the recent data~\cite{rdata} from BES and CMD--2. 
A conservative, `data--driven' fit~\cite{bp} gives
\be
\dahad = 0.02761 \pm 0.00036 \, ,
\label{adat}
\ee
corresponding to the `default' input of the blueband plot, 
whereas a more aggressive, `theory--driven' fit~\cite{mor}, 
corresponding to the `alternative' input of the blueband 
plot, gives 
\be
\dahad = 0.02738 \pm 0.00020 \, .
\label{athe}
\ee
There are many other determinations, as reviewed for example
in Ref.~\cite{jeg}, all consistent with the previous ones,
but with a tendency to be closer to Eq.~(\ref{adat}) than
to Eq.~(\ref{athe}). The second important uncertainty in the 
input parameters is the one associated with the experimental 
determination of the top quark mass from the CDF and D0 
experiments at Fermilab~\cite{top}:
\be
m_{top} = 174.3 \pm 5.1 {\rm \; GeV} \, .
\label{mtop}
\ee 

\begin{figure}[ht]
\epsfxsize200pt
\figurebox{120pt}{160pt}{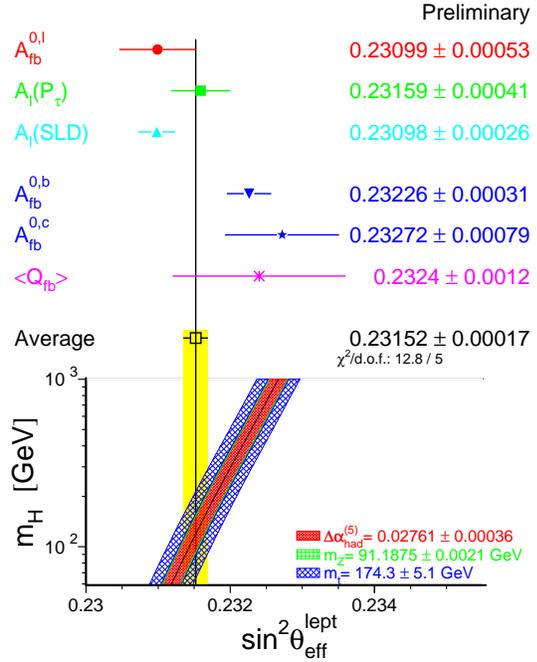}
\caption{Determination of $\s2t$ from the asymmetry 
measurements. The SM prediction is also shown, as a 
function of $m_H$, with the uncertainties from $\dahad$ 
and $m_t$ added linearly.}
\label{figs2t}
\end{figure}

\begin{figure}[ht]
\epsfxsize200pt
\figurebox{120pt}{160pt}{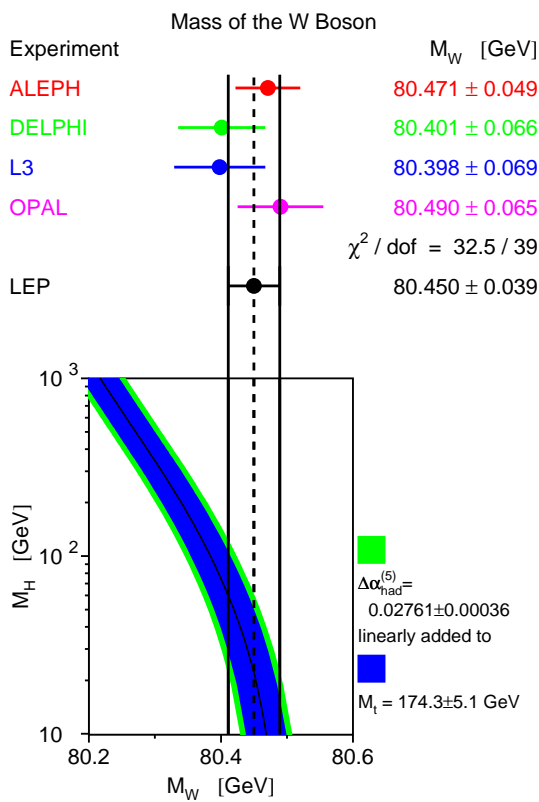}
\caption{The measurements of $m_W$ at LEP. The lower plot 
shows the SM prediction, as a function of $m_H$, with the 
uncertainties from $\dahad$ and $m_t$ added linearly.}
\label{figmw}
\end{figure}

The individual experimental determinations of $\s2t$ and $m_W$,
with the corresponding theoretical predictions and uncertainties,
as taken from Ref.~\cite{lepewwg}, are displayed in Figs.~\ref{figs2t} 
and \ref{figmw}.

A careful inspection of Figs.~\ref{fig:blueband}--\ref{figmw} 
reveals that the SM fit is not entirely a bed of roses, as
stressed, for example, in Ref.~\cite{cha} (on the basis of 
the data available in Winter 2001). First, the quality of the 
overall fit turns out to be acceptable but not exceptional, 
$\chi^2/dof = 22.9/15$, corresponding to a `probability' of $8.6 
\%$. The main reason for this can be seen from Fig.~\ref{figs2t}: 
there is a systematic tendency of the hadronic asymmetries ($b 
\ov{b}$ and $c \ov{c}$ forward--backward asymmetry plus quark 
charge asymmetry) to give larger values of $\s2t$ (and of $m_H$) 
than the leptonic asymmetries (forward--backward asymmetry, $\tau$ 
polarization asymmetry, SLD left--right asymmetry). The average 
of the hadronic determinations alone gives $\s2t(had) = 0.23230 \pm 
0.00029$, the average of the leptonic ones $\s2t(lep) = 0.23113 
\pm 0.00021$, corresponding to a discrepancy at the level of
$3.3 \sigma$. This effect was larger in Winter 2001: the change
is mostly due to a $-0.5\sigma$ shift of the $b \ov{b}$ 
forward--backward asymmetry, $A_{FB}^{0,b}$, after a new DELPHI 
analysis based on a neural network to tag the b--charge, and an 
improvement in the jet--charge measurement of the ALEPH analysis. 
The most precise hadronic determination comes indeed from 
$A_{FB}^{0,b}$, and has now a pull of $2.9 \sigma$ with respect 
to the central value of the global SM fit. Keeping in mind the
possibility of a statistical fluctuation, this can be viewed 
as a small problem either for the SM or for the experimental 
analyses. Radical modifications of the $Zb\ov{b}$ vertex appear 
unlikely, given the fact that $A_b$ from SLD and $R_b$ are 
well--behaved. Also, measuring $A_{FB}^{0,b}$ is a very delicate
experimental task, since flavor and charge of the b--quarks need
to be tagged simultaneously, with more complicated systematics
than in the measurement of $R_b$. On the other hand, all the 
experimental determinations of $m_W$ are in good agreement and
point to a light Higgs boson: those of Fig.~\ref{figmw} can be 
combined with the ones from the UA2, CDF and D0 experiments at 
$p \ov{p}$ colliders, giving $m_W = 80.454 \pm 0.060$, to produce 
a global world average $m_W = 80.451 \pm 0.033$. Notice that,
with the present errors, the main parametric uncertainty 
affecting the theoretical determination of $m_W$ is the one
coming from $m_t$, whereas $m_t$ and $\dahad$ give comparable
uncertainties in the theoretical determination of $\s2t$.

Given the small discrepancy between hadronic and leptonic
asymmetries, the exercise of looking at what happens, when
dropping the hadronic asymmetries from the fit, may not be 
entirely academical. The result is the following: the
quality of the SM fit improves, but the central value
of $m_H$ is pushed down, so that the consistency of the
SM fit to $m_H$ with the limits from direct searches
becomes marginal [with a significant residual dependence,
which should not be forgotten, on $\dahad$ and $m_t$].
Is there a SM crisis lurking around the corner? A prudent 
attitude before answering this question may be appropriate, 
taking into account that the final heavy--flavor analyses 
from LEP and improved determinations of $m_W$ and $m_t$ 
from the new Tevatron run will be available soon.
It may well be, however, that we must wait until the 
discovery (or the exclusion) of a light Higgs boson to
definitively settle the issue.

\subsection{The SM as an effective theory}\label{subsec:SMeff}

If we believed that the SM is the whole story, the talk 
could end here. However, we all know that the SM cannot 
be the ultimate theory of elementary particles,
valid at arbitrarily high energy scales, since it does not
contain a quantum theory of gravitational interactions and
some of its couplings are not asymptotically free. Thus,
the SM must be seen as an effective field theory, valid
up to some physical cut--off scale $\Lambda$, where new
physics must be introduced into the theory. On general grounds,
$\Lambda$ could be anywhere between the TeV scale and the
Planck scale, $M_P \equiv G_N^{-1/2}/\sqrt{8 \pi} \simeq
2.4 \times 10^{18} {\; \rm GeV}$, where $G_N$ is Newton's
constant, characterizing the observed gravitational interactions. 
Assuming that the SM correctly identifies the degrees of freedom 
at the weak scale (this may not be true, as will be discussed 
later, in the case of the Higgs field), we can write down the 
most general local Lagrangian compatible with the SM symmetries,
classifying the possible operators according to their 
physical dimension, and scaling all dimensionful couplings 
by appropriate powers of $\Lambda$. The resulting dimensionless 
coefficients are then to be interpreted as parameters, which 
can be either fitted to experimental data or (if we are able to 
do so) theoretically determined from the fundamental theory replacing
the SM at the scale $\Lambda$. Very schematically (and omitting
all coefficients and indices, as well as many theoretical 
subtleties):
 
\bea
{\cal L}_{eff}^{SM} & = & 
\left[ \Lambda^4 + \Lambda^2 \Phi^2 \right]
+ \left [ \left( D \Phi \right)^2 + \overline{\Psi} 
\not{\! \! D} \Psi \right.
\nn \\[4pt]
&&{} 
\left. + F^{\mu \nu} F_{\mu \nu} + F^{\mu \nu} 
\tilde{F}_{\mu \nu} + \overline{\Psi} \Psi \Phi 
+ \Phi^4 \right]
\nn \\[4pt]
&&{} 
+
\left[
{\overline{\Psi} \Psi \Phi^2 \over \Lambda}
+
{\overline{\Psi} \sigma^{\mu \nu} \Psi F_{\mu \nu} 
\over \Lambda}
+
{\overline{\Psi} \Psi \overline{\Psi} \Psi \over \Lambda^2}
\right.
\nn \\[4pt]
&&{} 
\left.
+
{ \Phi^2 F^{\mu \nu} F_{\mu \nu} \over \Lambda^2}
+
\ldots \right] \, ,
\label{leff}
\eea
where $\Psi$ stands for the generic quark or lepton field,
$\Phi$ for the SM Higgs field, $F$ for the field strength
of the SM gauge fields, and $D$ for the gauge--covariant 
derivative. The first bracket in Eq.~(\ref{leff}) contains
two terms, a cosmological constant term and a Higgs mass
term, that are proportional to positive powers of $\Lambda$,
and are at the origin of two infamous hierarchy problems.
The second bracket in Eq.~(\ref{leff}) contains operators 
with no power--like dependence on $\Lambda$, but only a milder,
logarithmic dependence, due to infrared renormalization effects 
between the cut--off scale $\Lambda$ and the weak scale. The 
last bracket in Eq.~(\ref{leff}) is the starting point of an
expansion in inverse powers of $\Lambda$, and contains operators
associated with rare processes, precision tests, neutrino masses, 
proton decay, \ldots. 

In this framework, an old question~\cite{cmmp} can be addressed 
in the light of present experimental data: given our knowledge 
of the top quark mass and of the bounds on the Higgs mass, can 
we put some firm bounds on the cut--off scale $\Lambda$? The 
qualitative aspects of the answer can be appreciated by 
remembering that, in the SM, the top and Higgs masses are 
associated with the largest Yukawa coupling $h_t$ and with 
the quartic Higgs self--coupling $\lambda$, respectively, via 
tree--level relations of the form $m_t \propto h_t v$ and 
$m_H^2 \propto \lambda v^2$, where $v$ is the vacuum expectation 
value of the Higgs field. Also, the scale-dependence of $\lambda$ 
is controlled by the renormalization group equation
\be
{d \lambda \over d \log Q} = {3 \over 16 \pi^2} 
\left( \lambda^2  + \lambda h_t^2 - h_t^4 \right) 
+ \ldots \, ,
\ee
where $Q$ is the renormalization scale and the dots stand for 
smaller one--loop contributions, controlled by the electroweak 
gauge couplings, and higher--order contributions. For any given 
values of $m_t$ and $\Lambda$, we can extract a `triviality'
upper bound on $m_H$ observing that, if $m_H$ is too large, 
$\lambda(Q)$ blows up at a scale $Q_0 < \Lambda$, developing 
a Landau pole. This leads to some well--known constraints, 
supported by more rigorous arguments and by lattice 
calculations: $m_H < 200$~GeV if $\Lambda \sim M_P$, $m_H 
< 600$~GeV if $\Lambda \sim 1$~TeV. Similarly, we can extract 
a `stability' lower bound on $m_H$ by observing that, if $m_H$ 
is too small, then $\lambda(Q)$ becomes negative at $Q_0 < 
\Lambda$, and another minimum of the SM potential develops 
at $\langle \phi \rangle \sim Q_0$.  

\begin{figure}[ht]
\epsfxsize200pt
\figurebox{120pt}{160pt}{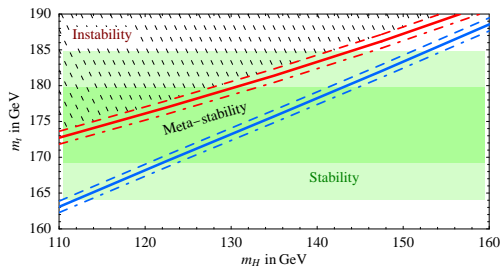}
\caption{Instability, meta--stability and stability regions
of the SM vacuum in the $(m_H,m_t)$ plane, for $\alpha_S (m_Z)
= 0.118$ (solid curves) $\pm 0.002$ (dashed and dash--dotted
curves). The shaded area indicates the experimental range
fro $m_t$, Eq.~(\ref{mtop}), at $1 \sigma$ (darker) and 
$2 \sigma$ (lighter).}
\label{irsa}
\end{figure}

\begin{figure}[ht]
\epsfxsize200pt
\figurebox{120pt}{160pt}{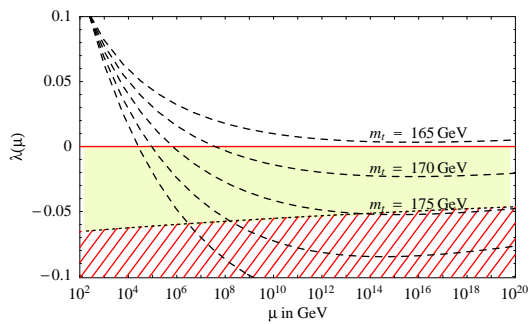}
\caption{Running of the quartic Higgs coupling $\lambda(\mu)$ for
$m_H=115$~GeV, $m_t=165,170,175,180,185$~GeV and $\alpha_S(m_Z)=
0.118$. Absolute stability [$\lambda(M_{weak})>0$] is still possible
if $m_t<166$~GeV. The hatched region is excluded by the meta--stability
bound.}
\label{irsb}
\end{figure}

Since the results of the previous subsection point to rather small 
values of $m_H$, the presently hot issue is the stability bound, 
recently revisited in~\cite{irs}. When implementing the stability
bound, three options are possible: 1) we can require absolute
stability, i.e. the correct electroweak vacuum must have lower 
energy than the `wrong' vacuum; 2) we can require stability with 
respect to high--temperature fluctuations in the cosmological
evolution of the early Universe; 3) we can require stability
with respect to quantum fluctuations at approximately zero 
temperature. The latter is the most conservative option, and
amounts to requiring that the lifetime of the correct
electroweak vacuum should be larger than the present age 
of the Universe, $T_U \sim 10^{10}$~yrs. The present
results are illustrated~\cite{irs} in Fig.~\ref{irsa}, 
where $\alpha_S(m_Z)=0.118 \pm 0.002$ and $\Lambda = M_P$ have been 
assumed, and in Fig.~\ref{irsb}, 
where $m_H=115$~GeV and $\alpha_S(m_Z) =0.118$ have been assumed. If 
we set $\Lambda \sim M_P$ (its maximum sensible value) and $m_H = 115
$~GeV (close to its minimum allowed value and to the location of the 
slight experimental effect discussed in subsection~\ref{subsec:direct}),
then option (1) leads to $m_t < (166 \pm 2)$~GeV. After a new 
complete one--loop calculation of the tunneling probability at
zero temperature, option (3) leads to $m_t < (175 \pm 2)$~GeV,
still in full agreement with the data. Therefore, it may be 
premature to claim evidence of new physics below $M_P$ from 
SM vacuum stability, even if we are at the border of the allowed 
region, a situation for which possible theoretical reasons have
been suggested~\cite{fnt}. 

\section{MSSM  (The Dogma?)}

In the SM effective Lagrangian of Eq.~\ref{leff}, the mass
term for the Higgs field has a quadratic dependence on the
cut--off scale $\Lambda$. When we try to extrapolate the SM
to scales much higher than the weak scale, this gives rise
to the infamous gauge hierarchy problem. The natural solution
to this problem is to introduce new physics close to the weak 
scale. The present best candidate for such new physics is the
Minimal Supersymmetric extension of the Standard Model (MSSM),
extensively discussed in another talk at this 
Conference~\cite{ellis}.

\subsection{Some virtues of the MSSM}

The main virtue of the MSSM is that, if the mass splittings
$\Delta m_{susy}$ that break supersymmetry (SUSY) are of the 
order of the weak scale $M_{weak} \sim 1$~TeV, then its 
cut--off scale can be naturally taken to be $\Lambda_{MSSM} 
= \Lambda_{susy}^2 / \Delta m_{susy}$, where $\Lambda_{susy}$ 
is the scale of spontaneous SUSY breaking. In hidden--sector 
supergravity models, where $\Lambda_{susy} \sim \sqrt{M_{weak} 
M_P}$, such cut--off scale can then be pushed very close to 
$M_P$ (this is not true if SUSY breaking occurs at lower scales, 
as in `gauge--mediated' models). 

\begin{figure}[ht]
\epsfxsize200pt
\figurebox{120pt}{160pt}{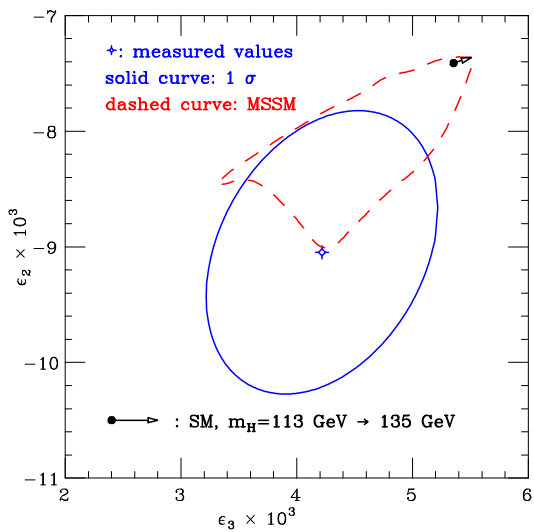}
\caption{Measured values (cross) of $\epsilon_3$ and $\epsilon_2$,
with their $1 \sigma$ region (solid ellipse), obtained from $m_W$,
$\Gamma_l$, $\s2t (lep)$ and $R_b$. The area inside the irregular
curve represents the MSSM prediction for $m_{\tilde{e}_L}$ between 
96 and 300~GeV, $m_{\chi^\pm}$ between 105 and 300~GeV, $|\mu|< 1
$~TeV, $\tan \beta = 10$, $m_{\tilde{e}_R}=1$~TeV and $m_A=1$~TeV.}
\label{duetre}
\end{figure}

\begin{figure}
\epsfxsize200pt
\figurebox{120pt}{160pt}{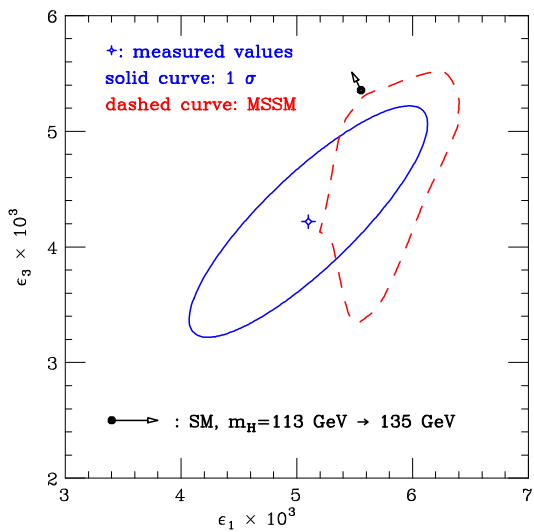}
\caption{The same as in Fig.~\ref{duetre} but for $\epsilon_1$
and $\epsilon_3$.}
\label{eps13}
\end{figure}

Another virtue of the MSSM is that, in contrast with other
possible solutions of the hierarchy problem, it is generically
as good as the SM in complying with electroweak precision
data. This is due to the fact that the soft SUSY--breaking
mass terms do not break the $SU(2)_L \times U(1)_Y$ gauge
symmetry. Indeed, it was recently observed~\cite{acggr} that, 
if sneutrinos and charged sleptons (and, to a lesser extent,
charginos and neutralinos) have masses close to their present
experimental bounds, then the MSSM may lead to an improved 
consistency between direct and indirect bounds on the Higgs 
mass, when the hadronic asymmetries are left out of the global 
fit. This result is illustrated in Figs.~\ref{duetre}
and \ref{eps13}, 
drawn in planes characterized by two of the three 
flavour--independent parameters $(\epsilon_1,
\epsilon_2, \epsilon_3)$ that are often used in
non--SM fits to precision data. We remind the reader
that $\epsilon_1$, related to Veltman's parameter
$\delta \rho$, is mainly controlled by $m_t$,
$\epsilon_2$ is particularly sensitive to $m_W$ and
$\epsilon_3$ is mainly controlled by $\s2t(lep)$.
Only the $W$ mass, the leptonic $Z$ width, the
leptonic asymmetries and the ratio $R_b$ have been
included in the fit. The elliptic solid contours represent 
the region allowed by the data at the $1 \sigma$
level. The irregular dashed contours enclose the typical 
MSSM predictions for a light spectrum. The fat dot with 
an arrow shows the SM prediction for a Higgs mass
varying between 113~GeV and 135~GeV. We can see
that, for a light MSSM spectrum, the agreement 
between data and theoretical predictions can improve.

Another important piece of indirect evidence in favour
of the MSSM is the fact that, when combined with a
condition on the grand unification of all gauge 
couplings and with the hypothesis of a `desert'
between the weak scale and the grand unification
scale, it leads~\cite{susygut} to one successful 
prediction for the gauge couplings at the weak scale. 
To gauge the significance of this success, we can perform
a simple--minded but illuminating exercise. We can
consider the one--loop renormalization group equation
for the running gauge coupling constants
\be
{d \alpha_A \over d \log Q} =
{b_A \over 2 \pi} \alpha_A^2 + 
\ldots \, ,
\;\;\;\;\;
(A=1,2,3) \, ,
\ee
where $b_A$ are the one--loop beta--function coefficients,
determined by the gauge quantum numbers of the particle spectrum
at the weak scale. If we are agnostic about the precise value of
the unified gauge coupling and of the grand unification scale,
but we assume the normalization of the $U(1)_Y$ gauge coupling
suggested by the simplest grand--unified models, we can perform 
a unification test by considering the only variable
controlling the prediction for the gauge couplings at the weak
scale, the ratio $B \equiv (b_3 - b_2)/(b_2 - b_1)$. The SM
value of this ratio is $B_{SM} \simeq 0.53$, its experimental
value is $B_{exp} \simeq 0.71$, and the MSSM value is $B_{MSSM}
\simeq 0.72$. A reasonable error estimate is $\Delta B \sim 0.03$,
completely dominated by the the fact that we do not know the details
of the MSSM spectrum and of the spectrum of the underlying theory
around the grand unification scale. This is an impressive success,
and it is difficult to believe that it is accidental and that we 
are being fooled by a malicious Nature and by theorists. Any other 
extension of the SM claiming to be better than the MSSM must face
this important phenomenological hint.

\subsection{The MSSM Higgs sector}\label{subsec:MSSMH}

\begin{figure*}
\epsfxsize30pc
\figurebox{16pc}{32pc}{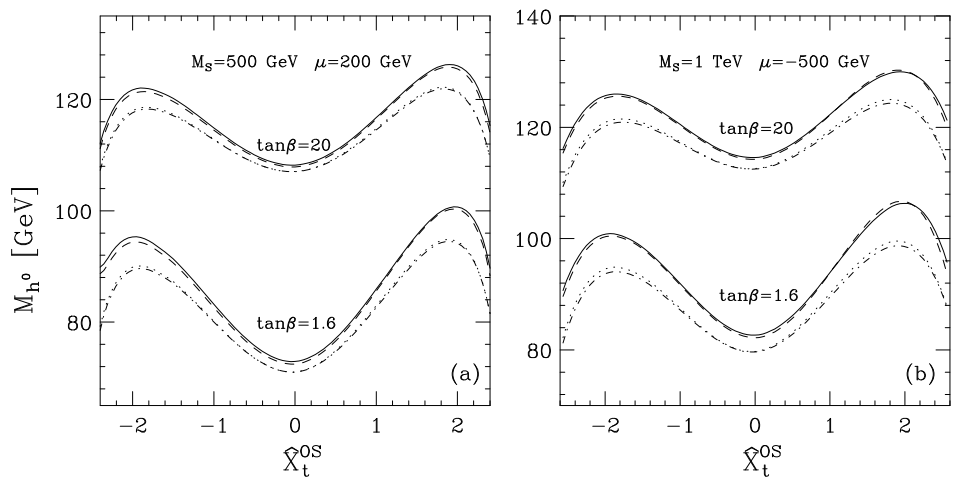}
\caption{The mass $m_h$ vs. the stop mixing parameter $\hat{X}_t^{OS}$,
for some representative values of the remaining MSSM parameters. The 
two--loop corrections are included either at ${\cal O} (\alpha_t 
\alpha_S)$ (lower lines) or at ${\cal O} (\alpha_t \alpha_S + 
\alpha_t^2)$ (upper lines). The fine structure corresponds to two 
different methods of implementing the corrections.}
\label{ez5}
\end{figure*}

If we take seriously the MSSM, then it is important to extract
its predictions for the Higgs sector. As is well known, the MSSM
Higgs sector contains two complex doublets, which after gauge
symmetry breaking give rise to five physical degrees of freedom,
three neutral $(h,H,A)$ and two charged $(H^\pm)$. The prediction
of SUSY is that the MSSM Higgs sector depends, at the tree level, 
only on known SM parameters and two more parameters, for example
$m_A$ and $\tan \beta \equiv v_2/v_1$. After including quantum 
corrections, the predictions of SUSY are not lost, but the 
dependences become more complicated and involve all the rest
of the MSSM spectrum, in particular the parameters of the
top--stop sector~\cite{shiggs}. An intense theoretical effort 
has been devoted over the last years to the precise computation
of the MSSM Higgs properties, and we are now at the stage
where the calculation of the most important two--loop 
corrections is being completed. When the top quark mass
will be known more precisely, these calculations will be
important for reliably comparing models of SUSY breaking 
with the available bounds on the spectrum. Of course,
the relevance of all this could increase further if and 
when SUSY particles and SUSY Higgs bosons will be found. 
So far, two--loop corrections to the neutral Higgs 
boson masses have been computed mostly in the limit 
of vanishing momentum on the external lines of the Higgs 
and gauge boson propagators. In this limit, analytical formulae 
at ${\cal O}(\alpha_t \alpha_S)$ are available, for
arbitrary values of the relevant MSSM parameters~\cite{asat},
and have been implemented in computer codes. As for the 
${\cal O}(\alpha_t^2)$ corrections, which can be of comparable
numerical importance, at the time of this Conference there are 
only partially analytic formulae~\cite{ez2} for $m_h$, valid in 
the limit $m_A \gg m_Z$. The general calculation of the ${\cal O}
(\alpha_t^2)$ corrections (in the zero--momentum limit) has been 
recently completed~\cite{bdsz} and agrees with Ref.~\cite{ez2} in 
the appropriate limit. The effects of the ${\cal O}(\alpha_t^2)$ 
corrections is illustrated in Fig.~\ref{ez5}, taken from 
Ref.~\cite{ez2}. 
We can see that these corrections can be sizeable, increasing $m_h$
by several GeV in the case of large mixing in the stop mass matrix.

\begin{figure}[ht]
\epsfxsize200pt
\figurebox{120pt}{160pt}{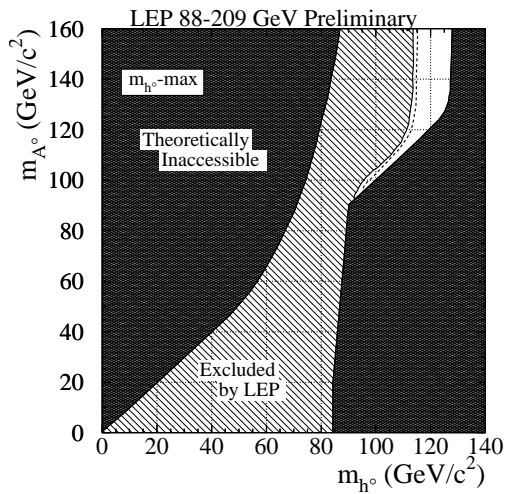}
\caption{The MSSM exclusion region in the $(m_h,m_A)$ plane,
for the `$m_h$--max' benchmark scenario. The central region 
is excluded by LEP searches, the lateral ones are
theoretically inaccessible in such a scenario.}
\label{mssm1}
\end{figure}

\begin{figure}
\epsfxsize200pt
\figurebox{120pt}{160pt}{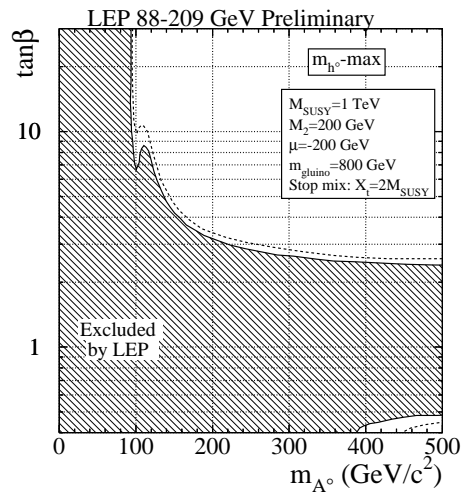}
\caption{The MSSM region of the $(m_A,\tan \beta)$ plane
excluded by LEP searches, for the `$m_h$--max' benchmark 
scenario.}
\label{mssm3}
\end{figure}

Armed with the relevant radiative corrections [the ${\cal O} 
(\alpha_t^2)$ ones have not yet been adequately implemented in 
the codes, but will presumably be included in the final LEP 
analyses], experimentalists have searched for direct signals of 
the MSSM Higgs bosons, as reviewed in another talk at this 
Conference~\cite{hanson} and described in more detail in 
Ref.~\cite{lepmssm}. The small Higgs signal
in the SM analysis has its counterpart in the MSSM
analysis: some excesses at the $\sim 2 \sigma$ level  
are reported both in the  $e^+ e^- \rightarrow h A$
channel, at $(m_h,m_A) \sim (83,83),(93,93)$~GeV,
and in the  $e^+ e^- \rightarrow hZ$ channel, for 
$m_h \sim 97,115$~GeV. The lower bounds on the MSSM
Higgs masses are notoriously difficult to illustrate,
due to their dependence on many parameters. Examples 
of exclusion plots are presented in Figs.~\ref{mssm1} 
and \ref{mssm3}, for a representative choice of MSSM parameters. 
In a `benchmark' case characterized by a large mixing in the stop mass 
matrix, which should lead to conservative bounds on $(m_h,m_A)$ and on
$\tan \beta$, the data have been interpreted~\cite{lepmssm} in terms 
of the following exclusion regions at $95 \%$ c.l.: $(m_h,m_A) < 
(91.0,91.9) {\rm \; GeV}$ and $\tan \beta < 2.4$. For small stop 
mixing, the limits are typically stronger.

There are other recent interesting studies of the MSSM Higgs
sector that would deserve to be discussed. There is just the 
time to briefly mention them, referring the reader to the 
corresponding papers. 

There is a new experimental analysis of the Tevatron data~\cite{charged},
on the search for $p \ov{p} \rightarrow b \ov{b} \varphi \rightarrow b 
\ov{b} b \ov{b}$ ($\varphi=h,H,A$). It leads to non--trivial 
constraints on the Higgs spectrum, in the region of the MSSM parameter 
space where the bottom Yukawa coupling is strongly enhanced with respect
to its SM value,  $\tan \beta > 40$--$50$.

Some recent theoretical studies~\cite{cphiggs} have considered the 
possibility of radiatively induced CP--violating effects in the Higgs 
sector, coming from explicit CP--violating phases in the squark--gluino 
sector, and have analyzed the resulting complications in the discussion 
of the MSSM Higgs searches. Other theoretical studies~\cite{adhsw}
have examined the implications of the experimental bounds on the
MSSM Higgses for different models of SUSY--breaking `mediation'.

\subsection{Some weak points of the MSSM}\label{subsec:MSSMw}

It would be misleading to end this section without mentioning 
that, besides its virtues, the MSSM has also, in our present 
view, a number of weak points. 

To begin with, the MSSM with
its soft SUSY breaking provides only an incomplete, technical 
solution of the hierarchy problem, since the overall mass scale 
of the soft terms is set `by hand'. These soft terms also 
introduce a very large number of free parameters into the model:
this problem is going to stay with us until when a standard 
model for spontaneous supersymmetry breaking will emerge
and/or SUSY particles will be found.

More seriously, after many years of experimental searches
at increasing energy scales, which explored a large part of 
the theoretically most appealing region from the point of 
view of the hierarchy problem, no direct experimental hint 
for the existence of the MSSM Higgs or SUSY particles has
been found.

Taking all this into account, we should not take the MSSM 
as a dogma for the new physics at the weak scale, but keep 
an open mind for the possible alternatives.

\section{Can~we~do~without~a~light~Higgs?
 \\ (The Heresy?)}

This part of the talk will touch an issue that often triggers 
heated discussions: can we do without a light Higgs? Some people 
view this as a heresy, some others almost take it for granted, 
so it is worth reviewing it, even if the state of affairs has 
not changed in an important way during the last year. Since, to 
the taste of most theorists, there is no satisfactory model 
without a light elementary Higgs, we may take an agnostic point 
of view and work at the level of an effective field theory.

The most drastic departure from the SM consists in getting 
rid of the elementary Higgs field, and having the electroweak
gauge symmetry $SU(2)_L \times U(1)_Y$ non--linearly realized
(so doing, of course, we define a non--renormalizable effective 
theory whose cut--off scale cannot be much above the TeV scale).
This approach has a long history, from the pioneering papers
of Ref.~\cite{nl1} to some recent phenomenological discussions
\cite{nl2} after the LEP and Tevatron data. In this approach, 
the effective Lagrangian is constructed from the Goldstone
bosons $w^a$ associated with electroweak symmetry breaking,
assembled into the group element $\Sigma = \exp ( 2 i  w^a 
\tau^a / v)$, where $v \simeq 256$~GeV. Concentrating on the
terms that can affect the $W$ and $Z$ propagators, thus playing
a major role in the discussion of electroweak precision tests,
we can write
\bea
{\cal L}_{eff} &  = & 
{v^2 \over 4} Tr \left( D_\mu \Sigma D^\mu \Sigma^\dagger \right)
\nn \\[4pt]
& + & \sum_i \tilde{c}_i {\widetilde{\cal O}}_i 
\left( \Sigma , \widetilde{\Lambda}, \ldots \right) \, ,
\label{leffnl}
\eea
where
\be
D_\mu \Sigma = \partial_\mu \Sigma 
+ i g W_\mu^a \tau^a \Sigma
-i g' \Sigma B_\mu^3 \tau^3 
\ee
is the covariant derivative. The first term in (\ref{leffnl})
describes the $W$ and $Z$ masses, as can be seen immediately 
in the unitary gauge $\Sigma=1$. The  higher--order operators
$\widetilde{\cal O}_i$ are scaled by appropriate powers of
the cut--off $\widetilde{\Lambda}$ of this Higgsless theory,
and are characterized by dimensionless coefficients 
$\tilde{c}_i$.

A less drastic approach consists in keeping the elementary
Higgs field $\phi$, so that $SU(2)_L \times U(1)_Y$ can be 
linearly realized, but in allowing the most general set of 
non--renormalizable operators compatible with the
electroweak gauge symmetry and with Poincar\'e invariance.
Also this approach has a long history, from the early paper
of Ref.~\cite{nr1} to other recent phenomenological discussions
\cite{nr2} after the LEP and Tevatron data. In this case,
the appropriate effective Lagrangian is
\be
{\cal L}_{eff} = 
{\cal L}_{SM} ( \phi )
+
\sum_i c_i  {\cal O}_i ( \phi, \Lambda, \ldots) \, ,
\ee
where $\Lambda$ is the cut--off and $c_i$ are the dimensionless
coefficients of the various operators ${\cal O}_i$.

In both approaches, the theoretical expressions for the two key 
pseudo--observables in the fits, $\s2t$ and $m_W$, differ from the 
SM ones. For given values of all the other parameters, in the SM 
they are just functions of $m_H$, with their leading dependences 
proportional to $\log (m_H/m_Z)$. In these new frameworks, the 
dependences become more complicated: 
\be
\log {m_H \over m_Z} \rightarrow
\left\{
\begin{array}{l}
\log {\widetilde{\Lambda} \over m_Z}
\\
\log {m_H \over m_Z}
\end{array}
\right.
+ 
K_{\theta,W}
\left( 
\begin{array}{c} 
\tilde{c}_i, \tilde{\Lambda}
\\
c_i, \Lambda
\end{array} \right)
\, .
\ee
In the Higgsless effective theory, the logarithmic dependence
on $\widetilde{\Lambda}/ m_Z$, generated by renormalization,
must be combined with finite contributions $K_\theta$ (for
$\s2t$) and $K_W$ (for $m_W$). A similar phenomenon occurs
in the effective theory with the Higgs field, with the only
difference that the logarithmic dependence is on $m_H/m_Z$. 
At the level of both effective theories, $K_{\theta,W}$  
depend on the cut--off and on the unknown dimensionless 
coefficients of the higher--dimensional operators, on which 
we can get reliable information only if we know about the 
underlying fundamental theory. With the present data, it is 
still possible to have $\widetilde{\Lambda} (m_H) \gg m_Z$ 
without excessive fine--tuning of the quantities $K_{\theta,W}$.

A more careful analysis, however, reveals the present advantage
of the light Higgs hypothesis. First, it must be said that, 
despite a lot of effort, so far there are no good candidates 
for the underlying theory that realizes the desired situation,
i.e. the phenomenologically correct magnitudes and signs of 
$K_{\theta}$ and $K_W$, without disrupting the predictions for
other observables, and avoiding `ad hoc' theoretical constructions.
Also, it can be immediately seen, in the linear realization, that 
there is an obvious correlation: if we increase $m_H$ we must
correspondingly decrease $\Lambda$, and tune the coefficients
$c_i$, to keep agreement with the data: then $m_H \sim m_Z$ 
and $\Lambda \gg m_Z$ looks as the most natural solution! 
There was a recent survey~\cite{pw} of models that may
evade the constraint of having a light Higgs with $m_H \sim m_Z$.
Three classes of models were identified, making reference to the
$(S,T)$ parameters, analogous to the $(\epsilon_3,\epsilon_1)$
parameters of Fig.~\ref{eps13}: (i) those in which new physics
can produce negative contributions to the $S$ parameter; (ii) 
those with a new vector bosons $Z'$ close to the weak scale; 
(iii) those in which new physics can produce positive 
contributions to the $T$ parameter. Without going to the details,
the important point is that all these models exhibit a rich 
phenomenology around the TeV scale, accessible to accelerators 
such as the Tevatron~\cite{kim}, the LHC~\cite{maiani} and a 
possible high--energy linear $e^+ e^-$ collider~\cite{heuer}. 

\section{New theoretical ideas \\ (Crackpot religions?)}

Electroweak symmetry breaking is a testing ground for new ideas 
that are restlessly explored by adventurous theorists, despite 
the fact that some of their most conservative colleagues may 
view them (in the words of Monty Python) as `crackpot religions'. 
It is appropriate to comment here on some ideas that have been 
very actively explored in the last years, focusing on the aspects 
that are most strictly related with Higgs physics.

The two big hierarchy problems of our present theories
are the cosmological constant problem and the gauge hierarchy 
problem, related with the two operators of dimension $d<4$
in the SM effective Lagrangian, conveniently rewritten as:
\be
{\cal L}_{eff}^{SM} = 
\Lambda_{cosm}^4
+
\Lambda_{weak}^2 \phi^2 
+
\ldots \, .
\label{lambdas}
\ee
We must explain why the scales of the vacuum energy and of the 
Higgs mass satisfy the phenomenological bounds $\Lambda_{cosm} 
\sim {\cal O}(10^{-3} {\; \rm eV})$ (as discussed in another
talk at this Conference~\cite{turner}) and $\Lambda_{weak} 
\sim {\cal O}(1 {\; \rm TeV})$, with the intriguing numerical
coincidence $\Lambda_{cosm} \sim \Lambda_{weak}^2 / M_P$. 

Are the two hierarchy problems related? It turns out that they 
are in supergravity and superstring theories, where supersymmetry
becomes a local symmetry and gravity is automatically included.
In these theories, formulating an acceptable model for SUSY
breaking is difficult, precisely because the problem of the weak 
scale (most probably linked to the scale of SUSY--breaking masses) 
and the problem of the cosmological constant scale must be 
addressed at once.

Many theorists feel that models formulated in more than
four space--time dimensions may offer unconventional
solutions to these problems and, perhaps, some exotic
phenomenology to be explored experimentally. Since these
models are the subject of another talk at this 
Conference~\cite{randall}, some comments related with 
electroweak symmetry breaking and the gauge hierarchy 
problem will be sufficient here (for other recent reviews 
and references on extra dimensions, see e.g. Ref.~\cite{extra}).


One of the most interesting features of models with extra 
dimensions is the fact that the hierarchy $M_{weak}/M_P$ 
can be linked with some geometrical object, in the simplest 
case a compactification radius $R$ characterizing the size 
of one or more compactified dimensions. Such a relation is 
strongly model--dependent. In toroidal compactifications, 
we can get power--like relations such as $(M_{weak}^2 / 
M_P^2) \propto R^{-n}$, where $n$ is an integer and the 
dimensionful proportionality coefficient is model--dependent. 
In `warped' compactifications, we can get exponential 
relations of the form $(M_{weak} / M_P) \sim \exp ( - M_P 
R )$. The gauge hierarchy problem is then reformulated 
in a very interesting way: it amounts to understanding the 
stability and the dynamical origin of the value of the radius 
$R$ that fits the phenomenological value of $M_{weak}$. 
There is no compelling idea so far in this direction, but 
some intriguing features are emerging and are at the center 
of an intense theoretical activity. 

Before describing some of the possibilities, it is worth 
mentioning that the problem of determining $R$ is analogous 
to the problem of understanding the stability and the dynamical 
origin of $\Delta m_{susy}$, the scale of SUSY--breaking mass 
splittings, in conventional, four--dimensional models of 
spontaneous SUSY breaking. The analogy becomes evident in those 
(higher--dimensional) superstring \cite{rohm,anto} and 
supergravity \cite{scsc} models where the radius $R$ does 
indeed control $\Delta m_{susy}$.

If there are symmetries of the higher--dimensional theory 
whose breaking is non--local in the extra dimensions,
symmetry--breaking quantities may be shielded from UV effects,
and determined by the infrared dynamics. As an example,
the field--dependent one--loop effective potential of 
some superstring~\cite{anto} (and field--theory~\cite{add}) 
compactifications does not contain positive powers of the 
string scale (cutoff scale $\Lambda$)
\be
V_1 (R,\phi) = R^{-4} + R^{-2} 
\phi^2 + \phi^4 + \ldots  \, , 
\ee
where all coefficients have been omitted and the dots stand for 
logarithmic corrections associated with the infrared running of 
the couplings. Starting from 
a higher--dimensional theory whose symmetries forbid a Higgs 
mass term (and ignoring the radius dynamics), $m_H$ and $v = 
\langle \phi \rangle$ are then calculable in terms of $R$, 
which may lead to one prediction. If there is no ultraviolet 
sensitivity, the program may be carried out at the field--theory
level~\cite{pred}, even if higher--dimensional field theories 
are in general badly non--renormalizable. However, if we lose 
the guidance of a consistent underlying superstring construction, 
there are many non--trivial consistency constraints to be satisfied, 
such as the absence of localized destabilizing divergences and 
anomalies, whose impact is being actively studied at the time of 
this Conference~\cite{after}.

We could be even more ambitious, and try to determine 
dynamically also $R$, if we could match the coefficient 
of the $R^{-4}$ term from the gravitational sector of the 
higher--dimensional theory. If all mass scales of the 
effective four--dimensional theory are controlled by the 
radius $R$, which in many compactifications is a 
classical flat direction because of a spontaneously broken 
scale invariance, then $R$ could be fixed at a finite 
non--zero value by a dimensional transmutation mechanism: 
an infrared fixed point for the $R$--dependent vacuum energy 
could arise from the interplay between the top--Yukawa and 
the gauge couplings~\cite{kpz}.

In the context of warped compactifications~\cite{randall}, an 
interesting mechanism for stabilizing $R$ at its desired value 
was suggested in Ref.~\cite{gw}. Since it involves the introduction 
of an explicit mass parameter, in a theory that had initially a 
classical scale invariance, it may be regarded as the analogous 
of soft breaking in the MSSM. This mechanism may be stable and 
related to a dimensional transmutation via the `holographic'
picture~\cite{hologw}.

Coming back to the main subject of the present talk, what features 
may emerge for Higgs phenomenology? It may be too early to tell. 
One possibility is the mixing between the Higgs boson(s), charged
under the electroweak gauge symmetry, and the spin--0 fields,
neutral under the electroweak gauge symmetry, that are associated
with the compactification radius (`radion') or, via supersymmetry,
with the Goldstone fermion of spontaneously broken supersymmetry 
(`sgoldstinos'). Such a mixing may lead to possible enhanced 
couplings of the Higgs to photons, gluons and invisible particles 
from the gravitational sector~\cite{radsgo}. This is a sufficiently 
good reason to keep an eye on non-standard Higgs searches~\cite{lepmssm} 
that do not assume the SM or MSSM production cross--sections and 
branching ratios.

In summary, the exploration of models with extra dimensions 
looks as a promising approach, in rapid development, with 
several controversial issues that have not been fully settled 
yet, and possible impact both on the theory and on the 
phenomenology of electroweak symmetry breaking.

\section{The ultimate answer \\ (The Universal Judgement)}

In the course of this talk we have been moving towards more 
and more speculative scenarios. How can we tell what is the
correct one? Fortunately, there is an impressive ongoing 
experimental program that will be able to tell us the answer.

\begin{figure}
\epsfxsize200pt
\figurebox{120pt}{160pt}{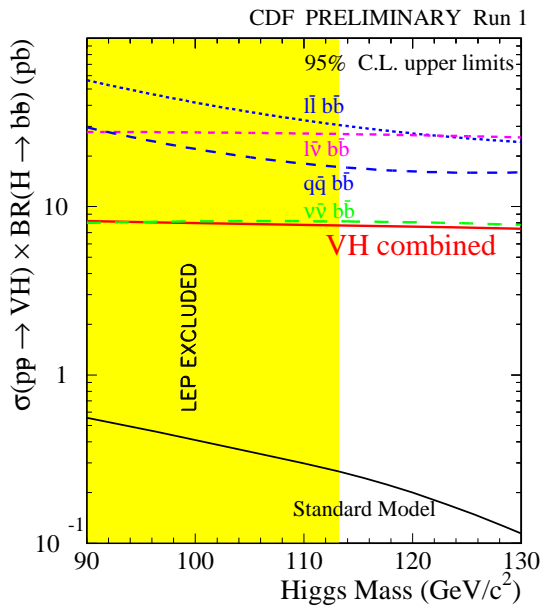}
\caption{Preliminary upper limits (at $95 \%$ c.l.) on 
production cross--sections times branching ratios, as 
functions of $m_H$, from Run I of the CDF experiment.}
\label{cdfhig}
\end{figure}

\begin{figure}
\epsfxsize200pt
\figurebox{120pt}{160pt}{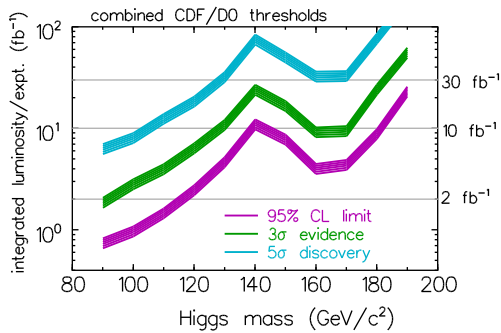}
\caption{The integrated luminosity required per experiment,
to either exclude a SM Higgs boson at $95 \%$~c.l. or discover
it at the $3 \sigma$ or $5 \sigma$ level, as a function of $m_H$.}
\label{tevreach}
\end{figure}

As discussed in another talk at this Conference~\cite{kim},
the Tevatron Higgs hunt is on its way. The present situation
of Higgs searches at the Tevatron is summarized in 
Fig.~\ref{cdfhig}, showing some preliminary results
from CDF~\cite{tevhiggs} (D0 had a slightly smaller 
sensitivity). With the luminosity and detectors of
Run I, Tevatron is still more than one order of
magnitude away from the sensitivity required by the
SM Higgs properties. However, as described in detail
in a dedicated study~\cite{tevgroup}, and summarized 
in Fig.~\ref{tevreach}, things will be different, and
very challenging, in the near future. For $m_H < 135$~GeV,
the focus of the present attention, CDF and D0 will
search for the SM Higgs boson considering its associated 
production with a weak gauge boson, $p \ov{p} \rightarrow 
V + (H \rightarrow b \ov{b})$ $(V=W^\pm,Z)$, and looking 
at a number of different final states: ($l \nu$)($b \ov{b}$),
($l^+ l^-$)($b \ov{b}$), ($\nu \ov{\nu}$)($b \ov{b}$).
Serious backgrounds are $V b \overline{b}$, $VV$, 
$t \overline{t}$, single top, and others. For $m_H >
135$~GeV, the channel $gg \rightarrow H \rightarrow W
W^{(*)}$ becomes accessible, and the useful final states
are ($l^\pm l^\pm j j$) and ($l^+ l^- \nu \ov{\nu}$).
(In this Section $l$ will always stand for $e$ or $\mu$.)
The results of Fig.~\ref{tevreach} are obtained by 
combining the statistical power of both experiments 
and all the channels mentioned above. The lower edge
of the bands is the calculated threshold; the bands 
extend upward from these nominal thresholds by $30 \%$
as an indication of the uncertainties in $b$--tagging
efficiencies, background rate, mass resolution, and other 
effects.

\begin{figure}
\epsfxsize200pt
\figurebox{120pt}{160pt}{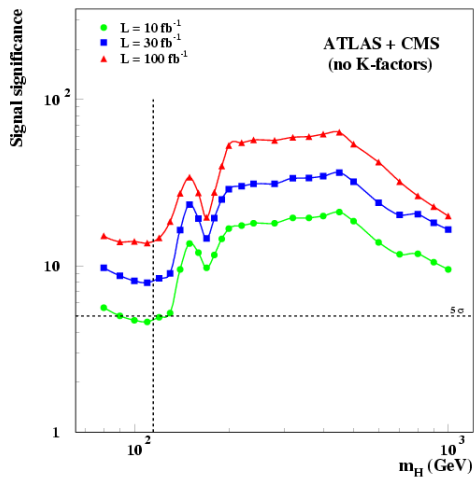}
\caption{Sensitivity for the discovery of a SM Higgs boson at
the LHC, as a function of $m_H$. The overall statistical 
significance, integrated over different channels, is 
plotted for three different integrated luminosities 
(10, 30 and 100~fb$^{-1}$), and assumes the combined 
statistical power of the ATLAS and CMS experiments.}
\label{lhc}
\end{figure}

The Higgs hunt will continue at the LHC (whose status is
summarized by another talk at this Conference~\cite{maiani}).
In the mass region $m_H > 130 {\; \rm GeV}$, the job of
the ATLAS and CMS experiments will be relatively easy, thanks
to the gold--plated channel $H  \rightarrow  Z Z^{(*)}   
\rightarrow  4 l^\pm$, with other channels as a backup 
for the mass regions with less statistics: $H  \rightarrow  
W W^{(*)}  \rightarrow  l \nu l \nu$ for $m_H \sim 2 \, m_W
\pm 30$~GeV, $H  \rightarrow ZZ  \rightarrow  l^+ l^- \nu \nu$
(and possibly $H  \rightarrow  W W  \rightarrow  l \nu j j $
or $H  \rightarrow ZZ  \rightarrow  l^+ l^- jj$) for $m_H >
600$~GeV. In the case of a light Higgs, $m_H < 130 {\; \rm 
GeV}$, various different signals are available. Earlier
studies have defined the strategies for signals such as 
inclusive $H \rightarrow \gamma \gamma$, $t \ov{t} + (H 
\rightarrow b \ov{b}, \, \gamma \gamma)$ and $V + (H 
\rightarrow b \ov{b}, \, \gamma \gamma)$. The combined 
discovery potential of the ATLAS and CMS experiments, for 
different integrated luminosities, is summarized~\cite{fg} 
in Fig.~\ref{lhc}.
During the last year, there was progress~\cite{kprz} 
in the study of the channel $q \overline{q} \rightarrow 
(WW \rightarrow  H) + j \, j$: exploiting the two tagged
forward jets, the background can be consistently reduced,
allowing the study of decay channels such as $H \rightarrow
W^{(*)} W^* \rightarrow l \nu l' \nu$, which may be a
discovery mode for $m_H \sim 115 {\, \rm GeV}$.

There are many other recent phenomenological studies on 
Higgs physics at high--energy colliders that would deserve
a detailed discussion. Time limitations just permit a brief
mention of some SM studies, with reference to the corresponding 
papers, leaving aside analogous studies for the MSSM and for
more exotic possibilities. Soft and virtual NNLO QCD 
corrections to $g g \rightarrow H + X$ have been
computed~\cite{gghx}. `Strong' weak effects at 
high--energies (Bloch--Nordsieck violations) were
studied~\cite{bn}: in particular, $(\alpha_W/\pi) \log^2 
(s/m_W^2)$ corrections to $\sigma (e^+ e^- \rightarrow 
{\rm hadrons})$ and an enhanced $m_H$ dependence in 
$W_L W_L \rightarrow {\rm hadrons}$. The cross--section
for Higgs $+$ 2 jets via gluon--gluon fusion was 
computed~\cite{hjj}. High-$p_T$ Higgs signals from 
$W W \rightarrow H \rightarrow b \ov{b}$ were
investigated~\cite{hpth}. The NLO QCD corrections to 
$p \ov{p} (pp) \rightarrow t \ov{t} H + X$ were
computed by two different groups~\cite{tthx}. 

\section{Conclusions}

In the presence of an experimental and theoretical puzzle,
as recalled by the title assigned to this talk by the 
Organizers, conclusions can only be tentative. 

It is clear that the search for the `Higgs boson' (or, 
more generally, for the dynamics underlying the spontaneous 
breaking of the electroweak gauge symmetry) is the 
main goal of high--energy physics in the present decade.

Direct searches and electroweak precision tests strongly 
constrain the possibilities: with the presently available 
information, the existence of at least one light Higgs 
boson with SM--like (or MSSM--like) properties looks like 
the best bet, but there is still room for the unexpected.

It is important to stress that, in all `natural' models, 
the Higgs boson is not alone: the accompanying physics 
may be even richer in implications (as, for example, in
the case of supersymmetry), and we must be prepared to 
fully explore the TeV scale.

While not very successful so far, the theoretical search
for plausible alternatives to the SM and the MSSM is 
worth pursuing, as confirmed by the many ongoing activities
along different directions, in particular extra dimensions.

The final (scientific) judgement is coming, and experiment 
will express it with the help of run II of the Tevatron, 
of the LHC, and hopefully more facilities to come \ldots 

In our quest for the fundamental laws of Nature, there is 
no substitute for the high-energy frontier!

\section*{Questions}
\begin{itemize}
\item[Q.] Bennie Ward, MPI and Univ. of Tennessee:\newline
There is the ultra--conservative view that we have a light Higgs, 
they will find it and there is nothing else. It is that way because 
God made it like that, unnatural or not. The theories in extra
dimensions you mention are non--renormalizable so, if you would 
find them, you are still left with their non--renormalizable 
artifacts. Why would you say one of these scenarios is better
than the other? 
\item[A.]
In view of naturalness arguments, the possibility of finding a
light Higgs and nothing else at the weak scale seems unlikely.
Rigorously, we cannot exclude that the gauge hierarchy problem 
is solved by mysterious infrared-ultraviolet connections that 
we are unable to understand with the tools of conventional 
quantum field theory. However, not even the cosmological
constant violates so far the naturalness criterion, since
gravitational interactions have been tested only up to energy
scales of the order of $10^{-3}$~eV, not far from the 
phenomenological value of $\Lambda_{cosm}$ in the normalization
of Eq.~(\ref{lambdas}). Coming to the second part of your
question, extra dimensions are just one out of many 
possibilities for the new physics at the Fermi scale. Their 
phenomenology may be described by an effective field theory, but 
the latter must eventually find an ultraviolet completion: this 
may be provided, for example, by an underlying superstring theory.
\end{itemize}

\section*{Acknowledgments}
The author would like to thank Riccardo Paramatti and Paolo Valente 
for precious technical help in the preparation of the talk, 
Giuseppe Degrassi for discussions on the content of section 1.2,
Juliet Lee--Franzini and Pietro Slavich for useful comments. This 
work was supported in part by the European Union under the contracts 
HPRN-CT-2000-00149 (Collider Physics) and HPRN-CT-2000-00148 (Across 
the Energy Frontier).
\end{document}